\documentclass[prb,twocolumn,showpacs,superscriptaddress]{revtex4}
\usepackage{amsfonts}
\usepackage{amsmath}
\usepackage{amssymb}
\usepackage{setspace}
\usepackage[dvips]{graphicx}
\usepackage{epsfig}
\usepackage{psfrag}
\usepackage{color}
\usepackage{latexsym}
\usepackage{comment}

\begin{document}
\title{Non-saturating magnetoresistance of inhomogeneous conductors: comparison of experiment and simulation}

\author{Jingshi Hu}
%\email{jingshi@uchicago.edu}
\affiliation{The James Franck Institute and Department of Physics,
The University of Chicago, Chicago, Illinois 60637, USA}

\author{Meera M. Parish}
%\email{mmp24@cam.ac.uk}
\affiliation{Cavendish Laboratory, JJ Thomson Avenue, Cambridge, CB3
0HE, United Kingdom}%
\affiliation{Department of Physics, Jadwin Hall, Princeton
University, Princeton, New Jersey 08544, USA}

\author{T. F. Rosenbaum}
%\email{t-rosenbaum@uchicago.edu }
\affiliation{The James Franck Institute and Department of Physics,
The University of Chicago, Chicago, Illinois 60637, USA}

%\date{\today, \now}
\date{\today}

\begin{abstract}
The silver chalcogenides provide a striking example of the benefits
of imperfection. Nanothreads of excess silver cause distortions in
the current flow that yield a linear and non-saturating transverse
magnetoresistance (MR). Associated with the large and positive MR is
a negative longitudinal MR. The longitudinal MR only occurs in the
three-dimensional limit and thereby permits the determination of a
characteristic length scale set by the spatial inhomogeneity. We
find that this fundamental inhomogeneity length can be as large as
ten microns. Systematic measurements of the diagonal and
off-diagonal components of the resistivity tensor in various sample
geometries show clear evidence of the distorted current paths
posited in theoretical simulations. We use a random resistor network
model to fit the linear MR, and expand it from two to three
dimensions to depict current distortions in the third (thickness)
dimension. When compared directly to experiments on
$\mathrm{Ag}_{2\pm\delta}\mathrm{Se}$ and
$\mathrm{Ag}_{2\pm\delta}\mathrm{Te}$, in magnetic fields up to 55
T, the model identifies conductivity fluctuations due to macroscopic
inhomogeneities as the underlying physical mechanism. It also
accounts reasonably quantitatively for the various components of the
resistivity tensor observed in the experiments.
\end{abstract}

\maketitle

\section{Introduction}

Macroscopic inhomogeneities were generally associated with a
diminution of galvanomagnetic effects, due to the reduction of
carrier mobility caused by impurity scattering. Classical effective
media theories~\cite{1,2,3,4,5,6,7,8} have shown, on the other hand,
that conductors normally exhibiting no change in resistance $R$ when
a magnetic field $H$ is applied, can possess a positive
magnetoresistance, $\Delta R/R \equiv (R(H)-R(0))/R(0)$, in the
presence of spatial fluctuations of the resistivity, provided that
the characteristic length scale of the inhomogeneity is much larger
than the mean free path. This induced magnetoresistance depends
generically on whether the magnetic field is oriented perpendicular
(transverse MR) or parallel (longitudinal MR) to the current
direction. In recent experiments, Solin and co-workers~\cite{9,10}
have harnessed this effect by crafting composite materials in which
current deflections around highly conducting inhomogeneities
significantly enhance the intrinsic transverse MR of
$\mathrm{Hg}_{1-x}\mathrm{Cd}_{x}\mathrm{Te}$. This designer
approach depends on the mismatch in the physical MR of the two
components and leads to a significant, quadratic magnetoresistance
that saturates by $H \sim$ 0.5 T.

In addition to geometric enhancements to the galvomagnetic response
of materials with intrinsic physical MR, there is the possibility of
designing materials and devices where the MR arises solely from the
macroscopic inhomogeneities.  In such a case, where the
disorder-induced component of the magnetoresistance can be separated
from the physical magnetoresistance, the functional form of the MR
not only can be tuned to be linear with magnetic field, but it is
possible to craft a response that continues to at least a MegaGauss.
Essential design considerations involve the pertinent inhomogeneity
length scale, which can involve both the dopant distribution and
polycrystalline structure, adjacency to a zero band gap condition to
amplify the effects of fluctuations in the carrier mobilities, and
the ease of growth and fabrication.

The silver chalcogenides stand out as favorable materials for the
investigation of large, linear and non-saturating MR because of
their simplicity and tunability. Perfectly stoichiometric
$\mathrm{Ag}_{2}\mathrm{Se}$ and $\mathrm{Ag}_{2}\mathrm{Te}$ are
non-magnetic, narrow-gap semiconductors whose electron and hole
bands cross at liquid nitrogen temperatures. They exhibit negligible
physical magnetoresistance~\cite{11}, as predicted from conventional
theories. By contrast, minute amounts of excess Ag or Te/Se $-$ at
levels as small as 1 part in 10,000 $-$ lead to a huge and linear
magnetoresistance over a broad temperature range, with no sign of
saturation up to 60 T~\cite{12,13,14,15,16,17}. In particular, the
unusual linearity extends deep into the low field regime with $H \ll
1/\mu$, where $\mu$ is the typical mobility of the material. Such
behavior shows no resemblance to conventional semiconductors, where
the magnetoresistance grows quadratically with field and reaches
saturation at fields typically of order 1~T. Therefore, it has been
argued that the observed magnetoresistance must be caused by the
inhomogeneous distribution of excess/deficient silver
ions~\cite{18,19}. The spatial inhomogeneities in the silver
chalcogenides result from the clustering of Ag ions at lattice
defects or grain boundaries when quenched from the high temperature
$\alpha$-phase to the semiconducting $\beta$-phase with lower Ag
solubility. They may take the form of highly conducting nanothreads
or lamellae along the grain boundaries of the polycrystalline
material~\cite{20}.

Abrikosov was the first to point out the essential role played by
inhomogeneities in the magnetoresistance observed in the silver
chalcogenides, emphasizing the importance of the zero band gap
condition. The phenomenon is quantum in origin and relies on a
combination of a linear dispersion and a single, partially-filled
Landau band~\cite{18,21}. This picture also has been applied to the
large, positive MR in epitaxial Bi films~\cite{22} and in suitably
doped InSb~\cite{23}. By contrast, Parish and Littlewood
(PL)~\cite{19,24} demonstrated that purely classical, geometric
effects of the inhomogeneities could be responsible for the
non-saturating magnetoresistance in
$\mathrm{Ag}_{2\pm\delta}\mathrm{Se}$ and
$\mathrm{Ag}_{2\pm\delta}\mathrm{Te}$. They introduced a
two-dimensional random resistor network model capable of simulating
macroscopic media with complicated boundaries. Their simulations of
strongly inhomogeneous semiconductors derive a linear MR from the
Hall resistance picked up from macroscopically distorted current
paths caused by variations in the stoichiometry. Moreover, when both
positive (hole-like) and negative (electron-like) values of the
mobility are sampled, as will occur at bank crossing, the functional
form of the MR is most linear (consistent with
experiments~\cite{17}). The magnetic field scale for the onset of
the linear response is then set by the width of the mobility
distribution. These results also agree qualitatively with subsequent
studies of two-component media involving exact results and effective
medium approximations~\cite{25,26,27}, which show that the
non-saturating transverse magnetoresistance will be maximal when the
high field Hall resistance crosses zero.

The exact distribution of inhomogeneities is difficult to resolve
via either scattering or imaging techniques for the relevant doping
levels in the silver chalcogenides, but the longitudinal MR has been
found to serve as an effective probe of the spatial extent of the
excess silver nanothreads~\cite{28}. The negative longitudinal MR
and, in particular, the emergence of an unexpected cross-term in the
resistivity tensor coupling the longitudinal and transverse
response, appears as a physical manifestation of distorted current
paths, with an attendant inhomogeneity length scale. Missing at
present, however, is the theoretical depiction of current
distortions in the thickness dimension, and the abnormal
longitudinal MR associated with it. The majority of theoretical
investigations on classical inhomogeneous conductors focus solely on
the transverse MR, the major exception being the study of isotropic
media with insulating inclusions, where the longitudinal MR is
substantially larger than the transverse MR~\cite{4,5}. Moreover,
the modeling has been restricted to transport properties in the
bulk, whereas the presence of boundaries is clearly important for
the unusual longitudinal MR of the silver chalcogenides.

In this article, we extend the PL random resistor model to three
dimensions in order to permit an explicit comparison to the
longitudinal MR results. We find that simple realizations of
macroscopic inhomogeneity represent a crucial ingredient for
understanding the full magnetoresistive response observed in the
experiments. The PL model allows one to simulate the various
components of the resistivity tensor, and generate visualizations of
the current distortion in the thickness dimension, results only
accessible through the computer simulations. Moreover, we test the
theory and fix its parameters via direct comparison to the
temperature and magnetic field dependence of the transverse MR.

The paper is organized as follows. Section II describes the sample
preparation and measurement techniques, while Sec. III reviews both
the 2D network model and the 3D expansion we use to visualize the
distorted current paths in the presence of a longitudinal magnetic
field. In Sec. IV, we compare the simulation results and
experimental data regarding the linear transverse magnetoresistance,
the transverse-longitudinal coupling $\rho_{xz}(H)$ and the negative
longitudinal magnetoresistance. We present our conclusions in Sec.
V.

\section{Experimental Methods}

Appropriately weighted amounts of high purity Ag and Te (99.999\%,
Alfa Aesar) were melted to created polycrystalline
$\mathrm{Ag}_{2\pm\delta}\mathrm{Te}$ at desired stoichiometries,
both silver rich ($n$-type, $\delta  > 0$) and silver deficient
($p$-type, $\delta < 0$).  The compound was ground and loaded into
outgassed, fused silica ampoules inside a helium glove box, then
baked at 50 K above the melting point (1170 K) to ensure complete
mixing. Slowly cooled specimens with typical dimensions (4.0
$\times$ 1.0 $\times$ 0.4) mm$^{3}$ were cut perpendicular to the
cylindrical axis to avoid possible dopant variations due to
temperature gradients inside the furnace.
$\mathrm{Ag}_{2\pm\delta}\mathrm{Se}$ polycrystals were prepared
similarly from stoichiometric $\mathrm{Ag}_{2}\mathrm{Se}$, and
appropriate weights of Ag or Se were added to reach the desired
compositions. Sample thicknesses from 400 $\mu$m to 15 $\mu$m were
obtained by mechanically shaving the same specimen with precise
depth control under the microscope.

We performed four-probe magnetotransport measurements using a
conventional ac bridge technique in the ohmic and
frequency-independent limits. Ultrasonically soldered InBi contacts
were placed on the top, bottom and sides of the specimens to
determine the $\rho_{xx}$, $\rho_{zz}$ and $\rho_{xz}$ components of
the resistivity tensor. The transverse and longitudinal MR was
determined by averaging values at positive and negative field
directions. The longitudinal MR is much smaller than the transverse
MR and thus requires attention to alignment of the magnetic field
and current directions (within $\pm 5^{\circ}$). We used a 14/16 T
superconducting magnet in a helium-3 cryostat for low field
measurements and the 60 T short-pulsed magnet at the National High
Magnetic Field Laboratory at Los Alamos for the high field response.
The latter provides 16 ms long field pulses from the discharge of a
1 MJ capacitor bank and, when combined with the NHMFL-designed
synchronous digital lock-in amplifier with response times much
shorter than commercial lock-ins, can provide low noise data during
each 16 ms shot.

\section{Theoretical Models}

For classical magnetotransport, where the mean free path of the
charge carriers is much less than the typical length scale of the
disorder, we can take Ohm's law to be valid locally in
space:
\begin{equation}
{\bf E}({\bf r}) = \hat{\rho}({\bf r}) \ {\bf j}({\bf r}).
\end{equation}
Here, $\mathbf{j}$ is the current density, $\mathbf{E}$ is the
electric field and $\hat{\rho}$ is the resistivity tensor. For
simplicity, we will assume that the inhomogeneities are isotropic
and that each point in space is characterized by a single charge
carrier, so that the resistivity tensor acquires the following form
in a magnetic field $\mathbf{H}=H\hat{z}$:
\begin{equation}
\hat\rho
\ = \ \rho \left (
\begin{array}{ccc}
1 & \beta & 0 \\
-\beta & 1 & 0 \\
 0 & 0 & 1
\end{array}
\right )
\end{equation}
Clearly, this tensor is dependent on just two parameters: the
carrier mobility $\mu$ (since the dimensionless variable $\beta =
\mu H$) and the scalar resistivity $\rho$. A realistic inhomogeneous
semiconductor will generally possess multiple charge carriers, but
their effect on $\hat{\rho}$ is typically small with
$\Delta\rho/\rho\ll 1$~\cite{29}.

In order to address the problem of strong inhomogeneities, PL
previously introduced a two-dimensional square network of
four-terminal resistor units~\cite{19,24}, which we will briefly
review here. As shown in Fig. 1(a), the resistor unit is taken to be
a homogeneous disk whose transport properties can be calculated
easily using Eq.~(1). Four terminals are required to take account of
the Hall voltage that develops perpendicular to the current
direction in a transverse magnetic field. The voltage differences
$\nu_{i}$ between terminals are linearly related to the currents
$\iota_{j}$ at each terminal via an impedance matrix $z_{ij}$, i.e.
$\nu_{i} = z_{ij} \ \iota_{j}$ . Like Eq.~(2), the impedance matrix
is completely determined by the mobility $\mu$ and an effective
scalar resistivity $\rho$. We can thus generate a square $N \times
N$ random network (Fig. 1(b)) by connecting up the disks using
Kirchoff's laws and then randomly varying $\mu$ and $\rho$ for each
disk. Specifically, we consider the distribution of $\mu$ to be
Gaussian with average $\langle\mu\rangle$ and width $\Delta\mu$,
while the distribution for $\rho$ must obviously satisfy $\rho\geq
0$ (e.g. we can take $\rho = \eta^2$, where $\eta$ has a Gaussian
distribution). Note that negative and positive values of $\mu$
correspond to electron-like and hole-like carriers, respectively,
and that the case where $\langle\mu\rangle = 0$ corresponds to an
equal number of electrons and holes since our chosen distribution is
symmetric about $\langle\mu\rangle = 0$. In order to determine the
network resistance $R$, we apply a potential difference $V$ across
the network in Fig.~1(b) and then sum up the currents along the left
(or right) edge, $I=\sum_i^N I_i^{L(R)}$, so that $R = V/I$. We
avoid boundary effects associated with perfectly conducting
electrodes on the left and right edges of the network~\cite{24} by
taking periodic boundary conditions for the currents, $I_i^L=I_i^R$.

\begin{figure}
\begin{center}
\includegraphics[width=0.8\linewidth]{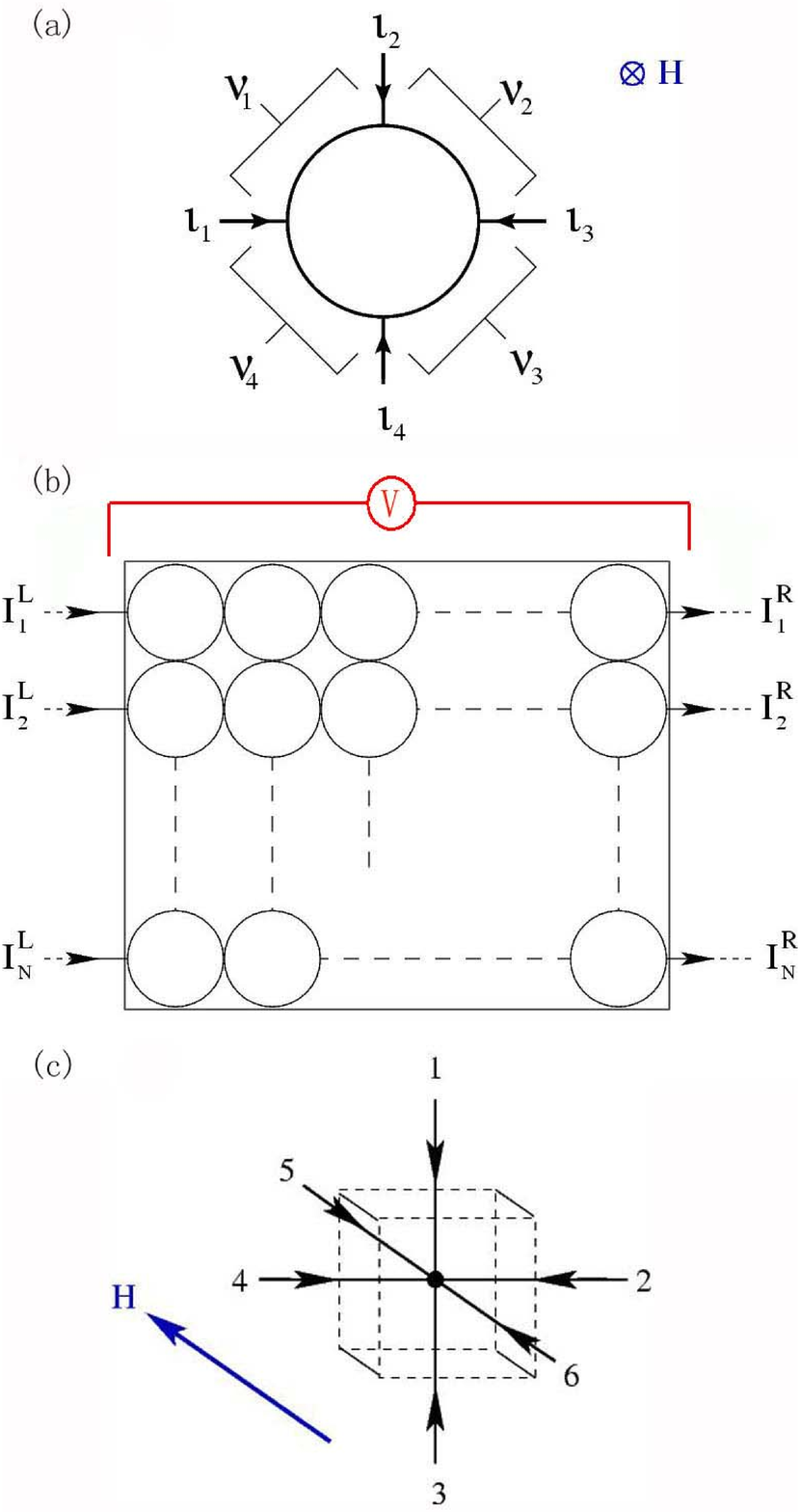}
\caption{Two-dimensional resistor network model and its extension to
three dimensions. The original 2D resistor unit (a) is assumed to be
a homogeneous disk with four equally spaced terminals, each of which
is associated with a current $\iota_{i}$ and a voltage difference
$\nu_{i}$ . The resultant $N \times N$ resistor network (b) has a
constant potential difference $V$ applied from left to right, and no
current can enter or exit the network at the top and bottom edges.
Unlike the previous resistor network model~\cite{19}, we assume
periodic boundary conditions for the currents on the left and right
edges $I_i^L=I_i^R$. In the three-dimensional version, the basic
unit is naturally a six-terminal element (c) of unit volume, but it
now represents a direct discretization of Eq.~(1) rather than any
particular geometry such as a sphere.}
\end{center}
\end{figure}

\begin{figure*}
\begin{center}
\includegraphics[width=0.8\linewidth]{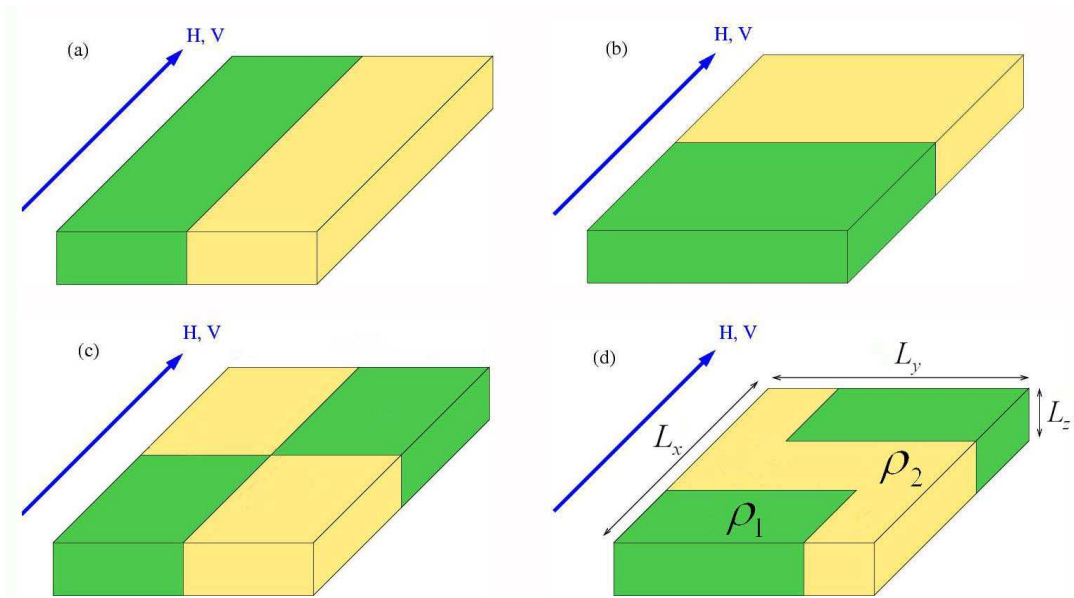}
\caption{Simple realizations of two-dimensional disorder consisting
of two different resistivities $\rho_1$ and $\rho_2$, where the
magnetic field $H$ is parallel to the applied voltage $V$. Patterns
(a) and (b) have no longitudinal magnetoresistance, while (c) has a
small, saturating one. The magnetoresistance of (d) can be tuned to
be large by setting $\rho_1/\rho_2 \gg 1$.}
\end{center}
\end{figure*}

In the limit $N \rightarrow \infty$, this model successfully
simulates the transverse MR of an inhomogeneous semiconductor.
However, in order to model the longitudinal MR, we must extend this
network to three dimensions, even in the case where the
inhomogeneity is purely two-dimensional. We should emphasize that
there are numerous subtleties associated with adding a third
dimension. First, the local resistivity tensor (2) is obviously not
isotropic like it is in 2D owing to the fact that there is now a
preferred direction along the magnetic field. Indeed, this property
is responsible for the standard phenomenon of current
jetting~\cite{29}. In addition, the resistivity in 3D contains the
dimensions of length (i.e. $R=\rho\frac{L_x}{L_y L_z}$) so it is not
scale invariant as it is in 2D. From a theoretical standpoint, the
extra dimension also makes it impossible to use duality
relations~\cite{7,8,25,30} or conformal mappings to obtain exact
results in 3D, although much progress has been made with effective
medium approximations for two-component systems~\cite{4,5,6,26}, and
by exploiting an analogy with advection-diffusion problems in the
continuous, weakly disordered case~\cite{2,3,31}.

The above considerations mean that there is no straightforward
generalization of the circular resistor unit to 3D. In fact,
deceptively trivial geometries like a sphere or a cylinder with its
axis perpendicular to the magnetic field have a non-saturating
magnetoresistance~\cite{32}. Instead, we must employ a
``brute-force'' approach by directly discretizing Eq.~(1) into
six-terminal elements as in Ref.~\onlinecite{33} and then
constructing 3D networks that simulate basic macroscopic
inhomogeneities. Referring to Fig.~1(c), we set a voltage and
current for each terminal such that each voltage is defined with
respect to the element center and incoming currents are defined as
positive. The impedance matrix relating voltages and currents is
then given by:
\begin{equation}
Z_{ij} = \frac{\rho}{2}(\delta_{ij} + \beta M_{ij})
\end{equation}
where
\begin{equation}
 \hat M = \frac{1}{2}
 \left(
 \begin{array}{cccccc}
  0 & -1 & 0 & 1 & 0 & 0 \\
  1 & 0 & -1 & 0 & 0 & 0 \\
  0 & 1 & 0 & -1 & 0 & 0 \\
  -1 & 0 & 1 & 0 & 0 & 0 \\
  0 & 0 & 0 & 0 & 0 & 0 \\
  0 & 0 & 0 & 0 & 0 & 0
  \end{array}
 \right)
\end{equation}
A major advantage of this model over effective medium approximations
is that we can include sample boundaries, a crucial ingredient for
understanding the observed negative longitudinal MR in the silver
chalcogenides~\cite{28}. In particular, we can simulate a four-probe
measurement, where the voltage probes placed on top of the sample
are separate from the electrodes at the ends of the sample that
supply the current (by contrast to the two-probe measurements
illustrated in Fig.~2).

The spirit of our investigation will be to construct simple
realizations of disorder that capture the salient features of
experiment, thereby gaining insight into an otherwise difficult
theoretical problem. We depict in Fig.~2 simple examples of possible
two-dimensional inhomogeneities with two different resistivities
$\rho_1$ and $\rho_2$. Note that there is no problem with using
perfectly conducting electrodes on the ends of the sample, since
unlike in a transverse field, there is no contact MR in a
longitudinal field. We see straightaway that the classical
longitudinal MR of Fig.~2(a) and (b) is zero since all the current
is parallel to the potential difference $V$ (or, equivalently, the
magnetic field $H$) and is thus insensitive to the Lorentz force.
However, once there is a non-zero current perpendicular to the
magnetic field, Hall voltages will build up in the $z$-direction,
which, in turn, can affect the current flow. The chessboard pattern
in Fig.~2(c) exhibits a small magnetoresistance for $\rho_1 \neq
\rho_2$ that eventually saturates once the current has found a path
across the sample that is parallel to $H$. The ideal pattern is one
where the current is forced to flow perpendicular to $H$ and
strongly experiences the Lorentz force as in Fig.~2(d) for
$\rho_1/\rho_2 \gg 1$. One may even regard this ``percolation''
pattern as representing a generic filamentary current path analogous
to those seen in the 2D network simulations~\cite{19,24}, although
in this case the current is filamentary even at zero magnetic field.
Surprisingly, differences in mobility $\mu$ between the two
components have hardly any effect on these simple patterns, so we
take $\mu$ to be uniform.

For the remainder of this article, we take Fig.~2(d) to be our basic
toy model for longitudinal magnetoresistance. We have computed the
magnetoresistance of different $L_x\times L_y\times L_z$ network
sizes and we find that it has converged for $L_x \geq 12$, $L_y \geq
16$ and $L_z \geq 24$ networks considered throughout.

\section{Results}

\subsection{Linear Magnetoresistance in a Transverse Magnetic Field}

In our initial studies of the silver chalcogenides we concentrated
on the galvanomagnetic behavior in a transverse magnetic field.
Semiclassically, the dependence of resistance on transverse field is
contained in the product $\omega\tau$, where $\omega$ is the
frequency at which the magnetic field causes the charge carriers to
sweep across the Fermi surface. Given a universal mean free time
$\tau \propto 1/n\rho$, where $n$ is the carrier density, the field
dependence of the transverse MR can be collapsed onto a universal
curve by expressing $\Delta R/R$ solely as a function of the
magnetic field, carrier density and the resistivity: $\Delta R/R =
f(\omega\tau) = f(H/n\rho)$.

\begin{figure}
\begin{center}
\includegraphics[width=0.9\linewidth]{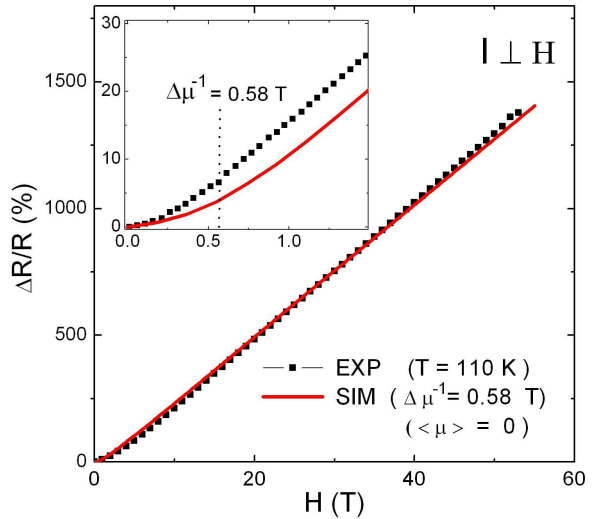}
\caption{Normalized transverse magnetoresistance, $\Delta R/R$, as a
function of magnetic field measured
at T = 110 K. The solid line is a theoretical fit from 20 $\times$
20 random resistor networks averaged over 10 realizations with
$\langle\mu\rangle=0$ and $\Delta\mu^{-1}$ = 0.58 T, consistent with
the PL hypothesis $\Delta\mu/\langle\mu\rangle > 1$. Inset: As seen
in a blow up of the low field region, the inverse of the mobility
distribution width, $\Delta\mu^{-1}$, sets the approximate scale for
the crossover to conventional quadratic behavior.}
\end{center}
\end{figure}

This functional dependence is known as Kohler's rule, obeyed by a
great number of metals and semiconductors, with a positive,
quadratic MR expected to saturate for $\omega\tau \approx 1$. By
contrast, in doped silver chalcogenides, the anomalous transport is
characterized by a large, linear magnetoresistance that shows no
sign of saturation at 60~T and extends down into the low field
regime ($H\ll 1/\mu$) where one generally expects $\Delta R/R
\propto H^2$. Moreover, the curves at different temperatures do not
coincide with the scaling form $\Delta R/R = f(\omega\tau)$, but are
shifted by a multiplicative factor that depends on
temperature~\cite{16}.

The PL model is able to account for the linearity at low fields by
showing that the linear, transverse MR of a heavily inhomogeneous
semiconductor crosses over to quadratic behavior at a field set by
the inverse of the mobility distribution width $\Delta\mu$, rather
than the average mobility $\langle\mu\rangle$, provided that
$\Delta\mu/\langle\mu\rangle > 1$ . In addition, the character of
the mobility distribution determines the magnitude of the linear
response: at sufficiently large fields, $\Delta R/R \propto
\Delta\mu H$ for strongly disordered semiconductors with
$\Delta\mu/\langle\mu\rangle > 1$ . Therefore, the galvanomagnetic
behavior of the silver chalcogenides naturally deviates from a
Kohler's plot, since its temperature dependence is no longer simply
tuned by the change in carrier mobilities, but also by the
disposition of the sample inhomogeneity.

\begin{figure*}
\begin{center}
\includegraphics[width=0.9\linewidth]{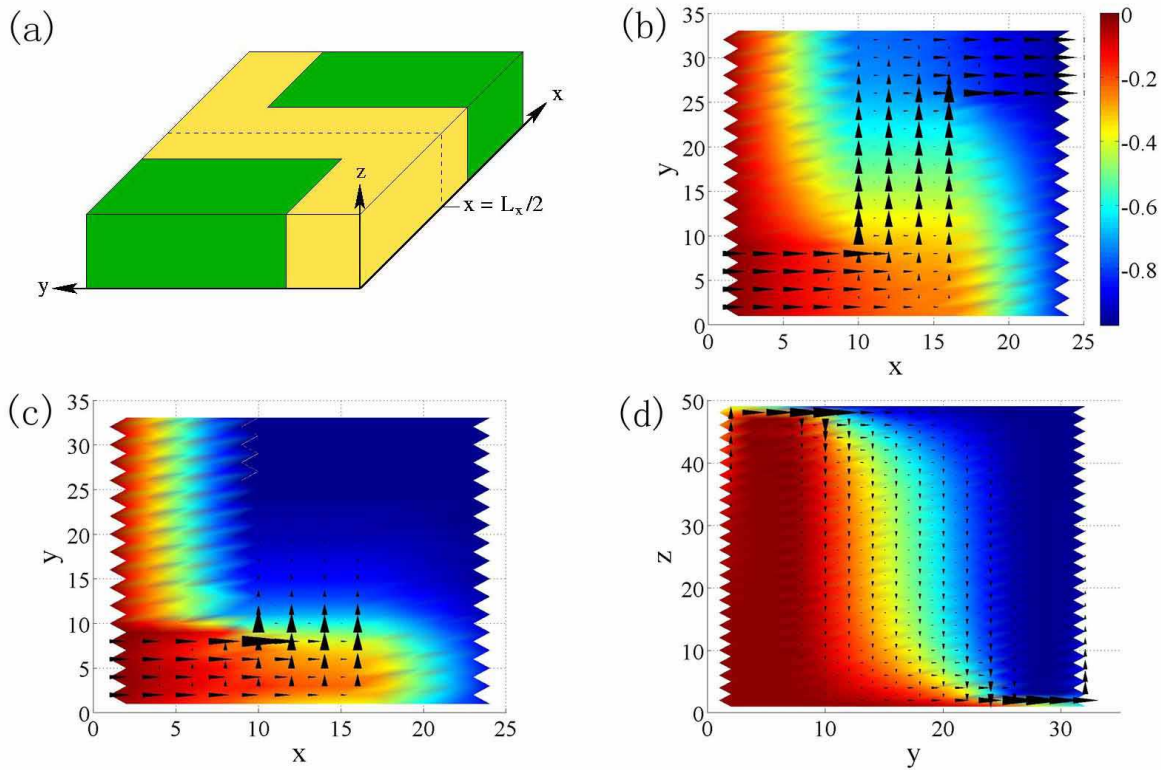}
\caption{Visualizations of currents and voltages for the
two-dimensional ``percolation'' pattern depicted in (a), taking
$\rho_1/\rho_2$ = 10,000 and using slightly smaller networks of 12
$\times$ 16 $\times$ 24 elements for better visibility. The currents
correspond to black arrows where the magnitude is given by the arrow
size, while the voltages are represented by different colors,
assuming the voltage drop across the sample is 1V. Panel (b) shows
the currents and voltages as viewed from the top in the absence of a
magnetic field, where the transport is independent of $z$. In a
large magnetic field $\beta$ = 10, the current develops a component
in the $z$-direction, resulting in a loss of current at the top
surface as depicted in (c). Panel (d) shows a slice at $x=L_x/2$
that reveals the behavior of the current in the $z$-direction.}
\end{center}
\end{figure*}

For a quantitative understanding of the temperature and field
dependence of the transverse MR in silver chalcogenides, it is
necessary to compare directly the theoretical predictions of the PL
model with experiment. Fits to the data will yield an estimate of
the mobility distribution in the silver chalcogenides and will test
whether the crossover from quadratic to linear MR is set by the
absolute value of the mobility $\mu$ or by the width of the mobility
distribution $\Delta\mu$. We plot in Fig.~3 the transverse MR in
$n$-type $\mathrm{Ag}_{2+\delta}\mathrm{Se}$ ($\delta$ = 0.0001) at
both low and high fields, fit to the random resistor network model.
We find that $\langle\mu\rangle=0$ yields the best fit, with
$\Delta\mu^{-1}$ = 0.58$\pm$0.05 T, confirming the expectation that
the silver chalcogenides are in the regime of
$\Delta\mu/\langle\mu\rangle \gg 1$, with a crossover field set
approximately by $\Delta\mu^{-1}$. While the value for
$\Delta\mu^{-1}$ is still slightly higher than the observed
crossover field, it is remarkably good given that we are considering
a simple random resistor network model with only two adjustable
parameters. Our toy model is oversimplified in many ways: it is
two-dimensional instead of three, it is restricted to a narrow range
of disorder types, and it neglects the contact effect between
network disks.
\begin{figure}
\begin{center}
\includegraphics[width=0.85\linewidth]{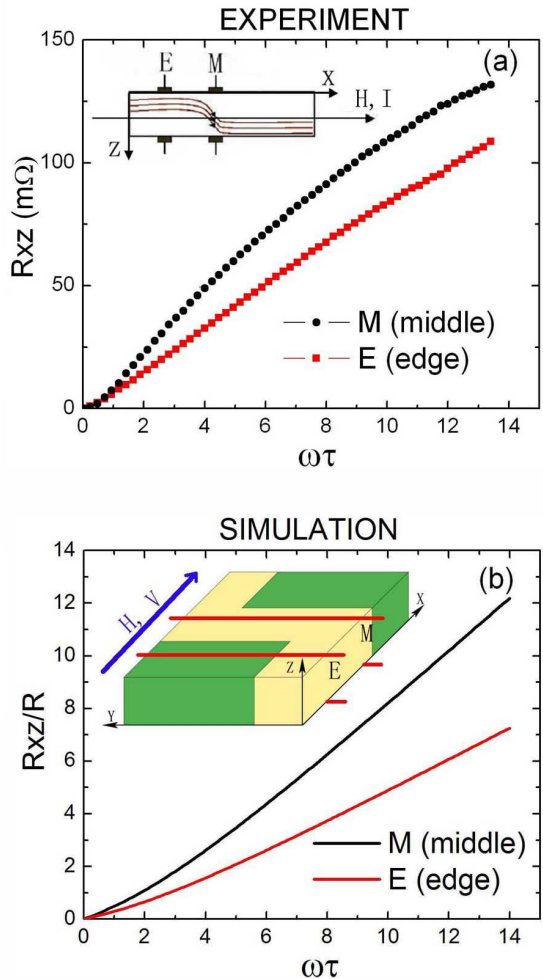}
\caption{(a) The voltage drop measured across the thickness of
$n$-type Ag$_2$Se, revealing distorted current flows as a direct
manifestation of sample inhomogeneities (depicted in the inset). A
conventional six-probe technique was used to compare signals from
the edge (E) and middle (M) of the specimen. (b) Numerically
simulated $R_{xz}(H)$ components of a 18 $\times$ 24 $\times$ 30
three-dimensional network (illustrated in the inset). The potential
on each side of the sample was averaged over the illustrative solid
lines.}
\end{center}
\end{figure}
Of particular interest is the fact that the electrons and holes are
present in equal proportions, as indicated by $\langle\mu\rangle=0$,
consistent with the claim that the magnetoresistance is most linear
when both positive and negative values of the mobility can be
sampled. For an inhomogeneous system, however, the averaged mobility
cannot be simply obtained from a measurement of the Hall resistance.
The experimental determination of the mobility distribution requires
a more precise characterization of the internal structure, as a real
material generally contains a small fraction of heavily doped
regions imbedded into a stoichiometric background with intrinsic
carrier concentration. Indeed, an unambiguous theoretical prediction
for the Hall resistance has proved elusive for the random resistor
network in this regime~\cite{24}.

\subsection{Current Distortions in a Longitudinal Field}

The MR in conventional semiconductors is caused by the curvature of
the carrier trajectories in a magnetic field. In a longitudinal
geometry, where the magnetic field lies in the same direction (the
$x$-direction) as the applied voltage, the carriers' momentum in the
$x$-direction is conserved for a uniform current flow, yielding a
nearly negligible longitudinal MR. The diagonal tensor components
$R_{xz}(H)$, representing transverse-longitudinal
couplings, should be zero due to the vector nature of the magnetic
field. When strong inhomogeneity exists, however, current
distortions in the $y$ and $z$ directions can significantly affect
the galvanomagnetic measurements.
%When strong inhomogeneity exists, however, it is necessary to
%consider the distorted current flows because the bending of
%equipotential lines can significantly affect the galvanomagnetic
%measurements.

While experiments cannot access local current patterns directly,
fortunately one can numerically simulate them. Our simple
two-dimensional percolation pattern in Fig.~4(a) provides an ideal
starting point for demonstrating the existence of field-induced
current distortions along the sample thickness. At zero field the
transport is independent of $z$, due to symmetry, and there is no
current in the $z$-direction (Fig.~4(b)). Once the sample is placed
in a large magnetic field ($\beta \gg 1$), a substantial current
flow develops in the $z$-direction. As shown in Figs.~4(c) and (d),
the current enters the sample along the top surface, flows along the
$z$-direction in the middle of the sample, and then exits along the
bottom surface. Note that the current is parallel to the
equipotentials at large fields, as expected, since the angle that
the current makes with the electric field is given by $\arctan
(\beta)$. The current pattern in Fig.~4(d) is reminiscent of that in
a rectangular homogeneous material with perfectly conducting
electrodes on the left and right boundaries [24]. There, the current
distortions were induced by a mismatch of the Hall resistivity
between the material and the electrodes. In this case, the mismatch
is generated from the change in the angle that the current spans
with the magnetic field, rather than arising from a change in the
properties of the charge carriers themselves.

Experimentally, we characterize the transport in the $z$-direction
of the silver chalcogenides by measuring the longitudinal-transverse
coupling $R_{xz}$, i.e. the voltage drop across the sample thickness
divided by the total current entering the sample. We plot in the
main panel of Fig.~5(a) the $R_{xz}$ tensor component for an n-type
$\mathrm{Ag_{2}Se}$ sample as a function of dimensionless field
$\beta=\omega\tau$ at T = 110 K. Here, a six-probe configuration was
used to compare the voltage drop at two different probe locations,
and allow for later correction of possible lead misalignments
($R_{xz}(H)-R_{xz}(0)\frac{\Delta R_{zz}(H)}{R_{zz}(0)}$).

The observed finite $R_{xz}$ provides clear evidence of the
predicted current distortion in the non-stoichiometric silver
chalcogenides system. Moreover, except for obvious changes in the
magnitude, the functional dependence on magnetic field of $R_{xz}$
appears to be rather insensitive to the location of voltage leads.
Fig.~5(b) provides the calculated $R_{xz}$ components of the
percolation pattern constructed from 12 $\times$ 16 $\times$ 30
elements, where the averaged potential on each side of the sample
was determined by sampling along the illustrated solid lines
(Fig.~5(b), inset). The simulated results show good consistency with
the experimental data, as demonstrated by an increasing $R_{xz}$
with increasing H without saturation up to 14 T ($\omega\tau \sim
14$). While both experiment and simulation have a similar dependence
on probe location, note that the theoretical $R_{xz}$ must decrease
as we approach the boundaries because the ideal electrodes do not
support a voltage drop.

The non-zero $R_{xz}$ observed in experiment and simulation
indicates that the inhomogeneities are macroscopic, with a length
scale that is appreciable compared to the sample dimensions. To see
this, it is important to stress that the sample is not invariant
with respect to a 180-degree rotation about the field axis,
indicating that the current paths favor a given direction in the
$y$-$z$ plane. Such symmetry breaking is not expected to occur in an
isotropic system that is effectively infinite, but finite sized
systems, such as our percolation pattern, can exhibit a particular
chirality. However, not all the salient features of experiment are
captured by our simple model. The theoretical $R_{xz}$ is purely
anti-symmetric with respect to magnetic field, while the
experimental $R_{xz}$ contains a substantial symmetric part. This is
not surprising given that the inhomogeneities in our model are only
two-dimensional. To generate a symmetric dependence of $R_{xz}$ on
field, we require current flow along the $z$-direction in the
absence of magnetic fields, and therefore three-dimensional
inhomogeneities, so that the $z$-component of the current is not
completely controlled by the field direction. Nonetheless, our
simple model serves to illustrate the principle of field-induced
current distortions across the sample thickness.

\subsection{Longitudinal Magnetoresistance}

The longitudinal MR provides the most striking manifestation of
sample inhomogeneities. As the distorted current now lies across the
thickness of the sample, bending the equipotential lines (Fig.~5(a),
inset), the voltage drop across the length of the sample will be
reduced significantly and a negative MR can emerge. We plot in
Fig.~6 the normalized longitudinal MR of $n$-type
$\mathrm{Ag_{2}Te}$ as a function of temperature and sample
thickness, and see a pronounced negative MR in the bulk sample (400
$\mu$m) for T $>$ 50 K. The maximum in the negative MR at T = 110 K
corresponds to the crossover from intrinsic to extrinsic
semiconducting behavior, where the distortions of current become the
strongest as the fluctuations in mobility can veer from positive to
negative. However, this effect is diminished for thinner samples,
such that there is no negative MR observed up to $H$ = 14 T for
thickness 20 $\mu$m. This demonstrates not only that current flow
across the sample thickness is crucial for the appearance of
negative MR, but that the critical thickness of 20 $\mu$m gives a
measure of the inhomogeneity length scale. The effects of spatial
randomness are suppressed completely if the size of the system
becomes comparable to the characteristic inhomogeneity length scale
in one dimension.

\begin{figure}
\begin{center}
\includegraphics[width=0.9\linewidth]{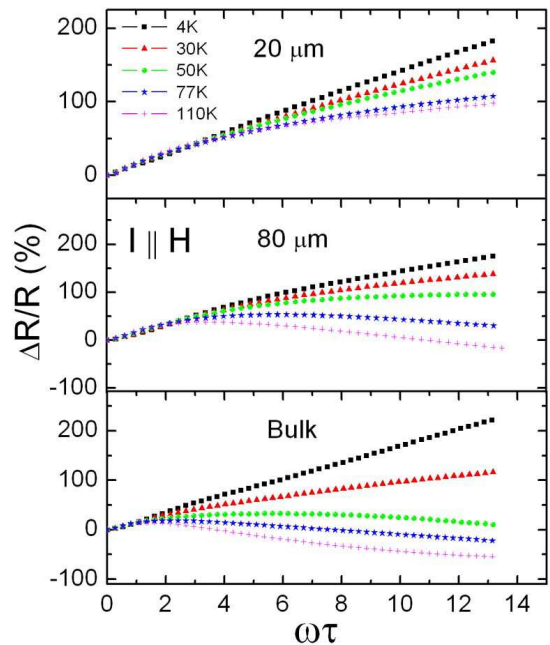}
\caption{Normalized longitudinal magnetoresistance vs. sample
thickness of $n$-type Ag$_2$Te. The negative component of the
magnetoresistance diminishes in thinner crystals, and the
inhomogeneity length scale is determined by the thickness where the
negative MR eventually disappears (20 $\mu$m).}
\end{center}
\end{figure}

We now compare the experimental results to simulations. For the
two-dimensional percolation pattern in Fig.~7(a) with
$\rho_1/\rho_2$ = 10,000, we know from Fig.~4(c) that there is a
flow of current away from the surface at high magnetic fields. Hence
a negative MR would be obtained from a four-probe measurement where
the voltage drop is measured by placing probes on the top surface of
the sample. However, the induced current distortions also increase
the actual resistance of the sample, and we find that this effect
dominates, resulting in the same positive linear magnetoresistance
as would be obtained from a two-probe measurement (Fig.~7(b)). This
is consistent with the experiments, where the magnetoresistance
becomes both positive and quasi-linear when we reduce the
inhomogeneities in the $z$-direction by thinning the sample. Similar
to the case of transverse MR, the linearity of the longitudinal MR
is derived from the Hall resistance contributed by distorted current
paths, namely by the currents in the $z$-direction of Fig.~4(d). The
magnitude of the longitudinal MR, however, is reduced when we
decrease $\rho_1/\rho_2$, as the magnetoresistance always saturates
at a field scale set by $\rho_1/\rho_2$. As expected, when we take
the limit $L_z \rightarrow 0$ in Fig.~2(d), we find that $\Delta R/R
\propto L_z$, so that there is no magnetoresistance in a truly
two-dimensional system.

\begin{figure*}
\begin{center}
\includegraphics[width=0.9\linewidth]{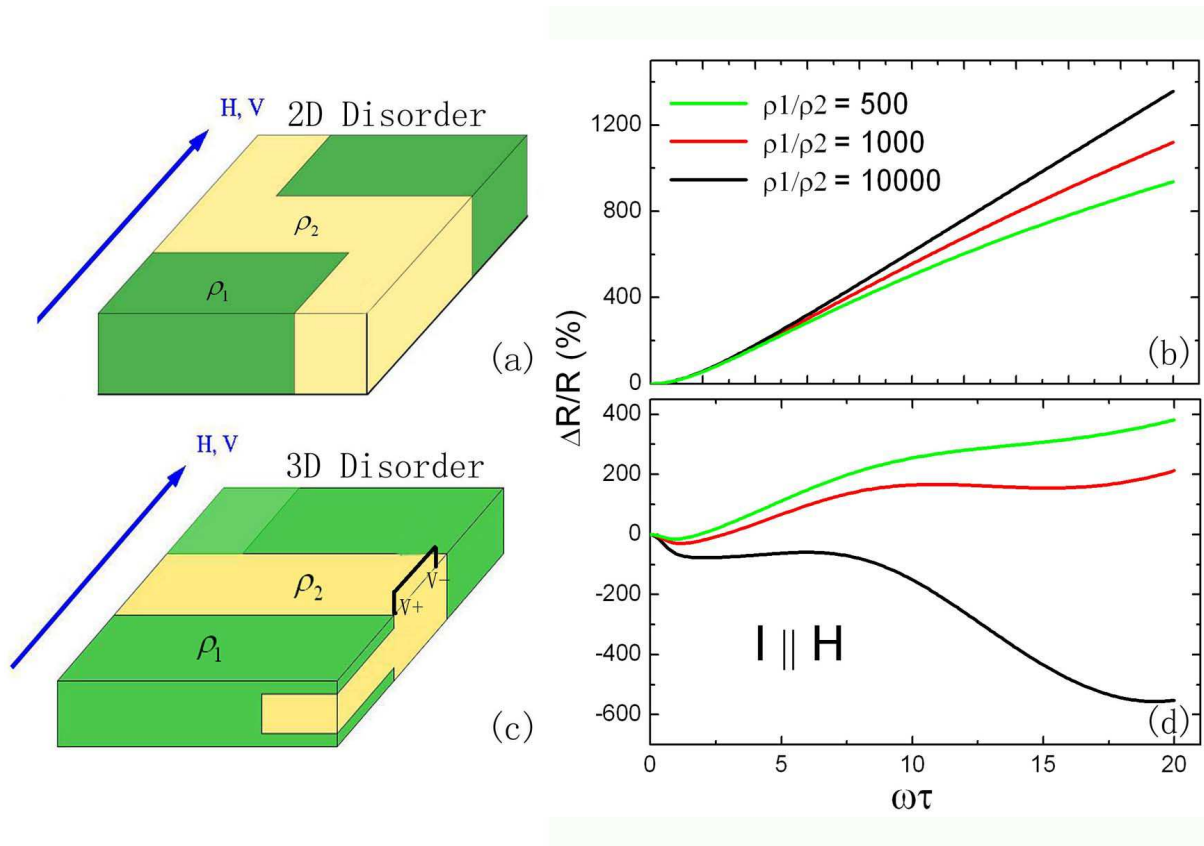}
\caption{Longitudinal magnetoresistance simulated using 18 $\times$
24 $\times$ 24 elements for the two-dimensional pattern and 18
$\times$ 24 $\times$ 40 elements for three-dimensional pattern. At a
sufficiently large resistivity ratio, corresponding to true
three-dimensional behavior, a negative longitudinal MR emerges.}
\end{center}
\end{figure*}

In order to obtain a negative magnetoresistance, one requires the
current to be depleted from the surface faster than the increase in
resistance. Since experiments imply that this scenario is possible
when there are inhomogeneities along the $z$-direction, we consider
a slight modification of our toy model to include three-dimensional
disorder, as depicted in Fig.~7(c). We perform a four-probe
measurement by averaging voltages along the solid lines in the
$y$-direction, and then take the component of the voltage difference
that is symmetric with respect to the magnetic field. Referring to
Fig.~7(d), we find a substantial negative magnetoresistance when
$\rho_1/\rho_2$  = 10,000, confirming that three-dimensional
inhomogeneities are responsible for the negative longitudinal MR
observed in the silver chalcogenides. It is of interest to note that
there is always an upturn in the negative magnetoresistance at high
fields for the simple models we considered, but it is unclear
whether this is generic for the infinite system.

\section{Conclusions}

In this paper, we have compared directly the galvanomagnetic
transport experiments on $\mathrm{Ag}_{2\pm\delta}\mathrm{Se}$ and
$\mathrm{Ag}_{2\pm\delta}\mathrm{Te}$ with theoretical simulations
of inhomogeneous conductors in both two and three dimensions. By
fitting the linear transverse MR with the PL 2D random resistor
network model, we have demonstrated that the silver chalcogenides
contain roughly equal proportions of electron- and hole-like
regions, and that the field scale for the onset of linearity is set
by the width of the mobility distribution rather than the average
mobility. Extending the model to three dimensions, we find that
simple realizations of macroscopic disorder are able to reproduce
the observed finite longitudinal-transverse coupling $R_{xz}$ and
the negative longitudinal MR in the silver chalcogenides. Moreover,
visualizations of the current distortions across the thickness
dimension illustrate the depletion of current at the sample surface
that is crucial for the phenomenon. Future work is required to
extend the theoretical results to the infinite system and to
characterize precisely the types of disorder which give rise to the
biggest magnetoresistive response.

\begin{acknowledgments}
We are grateful to P.~B.~Littlewood and M.-L.~Saboungi for
illuminating discussions, and to J.~B.~Betts for the help with the
high-field measurements at NHMFL. The work at the University of
Chicago was supported by US Department of Energy Grant No.
DE-FG02-99ER45789. J.~H. acknowledges support from the Consortium
for Nanoscience Research.
\end{acknowledgments}

\end{document}